\documentclass[cameraready]{Interspeech}

\title{Privacy-Preserving End-to-End Full-Duplex Speech Dialogue Models}

\author[affiliation={1,2}]{Nikita}{Kuzmin}
\author[affiliation={3,4}]{Tao}{Zhong}
\author[affiliation={3},correspondingauthor]{Jiajun}{Deng}
\author[affiliation={3}]{Yingke}{Zhu}
\author[affiliation={3}]{Tristan}{Tsoi}
\author[affiliation={3}]{Tianxiang}{Cao}
\renewcommand{\authorsep}{,\\}
\author[affiliation={3}]{Simon}{Lui}
\author[affiliation={5}]{Kong Aik}{Lee}
\author[affiliation={1}]{Eng Siong}{Chng}

\address{
  $^1$Nanyang Technological University, Singapore \quad
  $^2$A*STAR, Singapore \\
  $^3$Huawei Leibniz Research Center \quad
  $^4$The Chinese University of Hong Kong \\
  $^5$The Hong Kong Polytechnic University
}
\email{s220028@e.ntu.edu.sg}


\usepackage{booktabs}
\usepackage{multirow}
\usepackage{siunitx}
\usepackage{amsmath}
\usepackage{colortbl}
\definecolor{anonrow}{gray}{0.93}
\newcommand{\gc}{\cellcolor{anonrow}}
\usepackage{pgfplots}
\pgfplotsset{compat=1.18}
\usepgfplotslibrary{fillbetween}

\definecolor{tolblue}{RGB}{68,119,170}
\definecolor{tolred}{RGB}{238,102,119}
\definecolor{tolgreen}{RGB}{34,136,51}
\definecolor{tolorange}{RGB}{238,119,51}
\definecolor{tolpurple}{RGB}{170,51,119}
\definecolor{tolcyan}{RGB}{68,187,153}

\definecolor{dangerDark}{RGB}{178,24,43}
\definecolor{dangerMid}{RGB}{214,96,77}
\definecolor{dangerLight}{RGB}{204,102,51}
\definecolor{safeDark}{RGB}{5,113,100}
\definecolor{safeMid}{RGB}{40,152,62}
\definecolor{safeLight}{RGB}{27,158,119}

\begin{document}

\maketitle

\begin{abstract}
End-to-end full-duplex speech models feed user audio through an always-on LLM
backbone, yet the speaker privacy implications of their hidden representations
remain unexamined.
Following the VoicePrivacy 2024 protocol with a lazy-informed
attacker, we show that the hidden states of SALM-Duplex and Moshi leak
substantial speaker identity across all transformer layers.
Layer-wise and turn-wise analyses reveal that leakage persists across all
layers, with SALM-Duplex showing stronger leakage in early layers while Moshi
leaks uniformly, and that Linkability rises sharply within the first few turns.
We propose two streaming anonymization setups using Stream-Voice-Anon: a
waveform-level front-end (Anon-W2W) and a feature-domain replacement
(Anon-W2F).
Anon-W2F raises EER by over 3.5$\times$ relative to the discrete encoder
baseline (11.2\%$\rightarrow$41.0\%), approaching the 50\% random-chance ceiling, while
Anon-W2W retains 78--93\% of baseline sBERT across setups with
sub-second response latency (FRL under 0.8\,s).%
\footnote{Demo page: \url{https://anonymous-569230593.github.io/}}
\end{abstract}

\section{Introduction}

End-to-end (E2E) full-duplex speech dialogue systems represent a fundamental shift
from turn-taking interaction to always-on, simultaneous listening and speaking.
Models such as SALM-Duplex~\cite{salmduplex2025}, Moshi~\cite{defossez2024moshi},
and SyncLLM~\cite{emnlp2024synclm} route raw user audio continuously through a
decoder-only LLM backbone, computing hidden state representations at every
transformer layer throughout the entire conversation.
Unlike cascaded systems where user and agent streams are processed separately,
the LLM core of these full-duplex systems maintains a persistent internal state
over the user's speech stream, capturing their voice, speaking style, and identity.

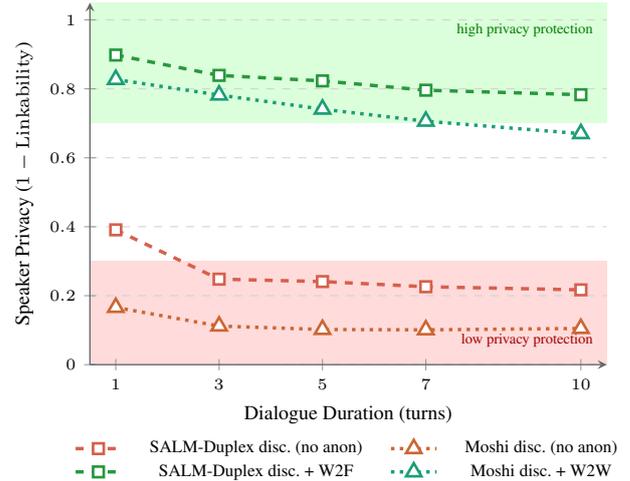
\begin{figure}[t]
  \centering
  \begin{tikzpicture}
  \begin{axis}[
    width=0.85\columnwidth, height=0.60\columnwidth,
    scale only axis,
    xlabel={Dialogue Duration (turns)},
    ylabel={Speaker Privacy ($1 - \mathrm{Linkability}$)},
    xlabel style={font=\fontsize{7.5}{8.5}\selectfont},
    ylabel style={font=\fontsize{7.5}{8.5}\selectfont},
    xmin=0.5, xmax=10.5, ymin=0, ymax=1.05,
    xtick={1,3,5,7,10},
    ytick={0,0.2,0.4,0.6,0.8,1.0},
    tick label style={font=\fontsize{6.5}{7.5}\selectfont},
    axis lines=left, axis line style={line width=0.5pt, black!60},
    tick style={black!40, line width=0.4pt},
    ymajorgrids=true, xmajorgrids=false,
    grid style={draw=black!18, line width=0.35pt, dashed},
    legend style={at={(0.5,-0.18)}, anchor=north, legend columns=2,
      draw=none, fill=none,
      font=\fontsize{6.5}{7.5}\selectfont,
      column sep=8pt, row sep=-1pt},
    set layers,
  ]
  \begin{pgfonlayer}{axis background}
    \fill[red!14]   (axis cs:0.5,0)    rectangle (axis cs:10.5,0.3);
    \fill[green!14] (axis cs:0.5,0.7)  rectangle (axis cs:10.5,1.05);
  \end{pgfonlayer}
  \node[font=\fontsize{5.5}{6.5}\selectfont, text=red!65!black,   anchor=east]
    at (axis cs:10.4,0.07) {low privacy protection};
  \node[font=\fontsize{5.5}{6.5}\selectfont, text=green!50!black, anchor=east]
    at (axis cs:10.4,0.97) {high privacy protection};

  \addplot[color=dangerMid, line width=1.3pt, dashed,
    mark=square*, mark size=2.0pt,
    mark options={fill=white,draw=dangerMid,line width=1.0pt,solid}]
    coordinates {(1,0.391)(3,0.248)(5,0.241)(7,0.226)(10,0.217)};
    \addlegendentry{SALM-Duplex disc.\ (no anon)}
  \addplot[color=dangerLight, line width=1.3pt, dotted,
    mark=triangle*, mark size=3.3pt,
    mark options={fill=white,draw=dangerLight,line width=1.0pt,solid}]
    coordinates {(1,0.166)(3,0.112)(5,0.102)(7,0.101)(10,0.105)};
    \addlegendentry{Moshi disc.\ (no anon)}

  \addplot[color=safeMid, line width=1.3pt, dashed,
    mark=square*, mark size=2.0pt,
    mark options={fill=white,draw=safeMid,line width=1.0pt,solid}]
    coordinates {(1,0.898)(3,0.839)(5,0.823)(7,0.796)(10,0.783)};
    \addlegendentry{SALM-Duplex disc.\ + W2F}
  \addplot[color=safeLight, line width=1.3pt, dotted,
    mark=triangle*, mark size=3.3pt,
    mark options={fill=white,draw=safeLight,line width=1.0pt,solid}]
    coordinates {(1,0.827)(3,0.782)(5,0.741)(7,0.706)(10,0.670)};
    \addlegendentry{Moshi disc.\ + W2W}

  \end{axis}
  \end{tikzpicture}
  \caption{Speaker privacy ($1 - \mathrm{Linkability}$) vs.\ dialogue
    turn count. Only discrete encoder variants are shown; continuous encoder
    Linkability is omitted for clarity.
    Red lines: no anonymization;
    green lines: post-anonymization.
    Without anonymization, both systems drop into the low-privacy zone within
    a few turns; anonymization lifts privacy into the protected zone,
    though Moshi + W2W shows gradual degradation over dialogue length.}
  \label{fig:turnwise_link}
\end{figure}

This creates a privacy exposure that remains largely unexplored.
Under GDPR and similar regulations, the mere presence of identifiable speaker
information in model representations constitutes a compliance risk, regardless
of whether external access is exploited, making it essential to audit and
mitigate such leakage proactively.
Probing studies on self-supervised speech models (e.g., wav2vec 2.0, HuBERT,
WavLM) have established that their hidden representations strongly encode
speaker identity~\cite{pasad2021layerwise, yang2021superb, chiu2025probing}.
In the text domain, prior work has shown that LLM hidden states can leak
demographic and content-level user attributes~\cite{staab2024beyond}, and
membership inference probes have exposed privacy risks in masked language
models~\cite{mireshghallah2022quantifying}.
Our focus is distinct: we investigate \emph{speaker identity} leakage, the
persistent encoding of \emph{who is speaking}, which enables re-identification
regardless of conversational content.
Yet, to the best of our knowledge, no prior work has examined whether these always-on LLM hidden states retain sufficient speaker identity to enable re-identification.

\begin{figure*}[t]
  \centering
  \includegraphics[width=\textwidth]{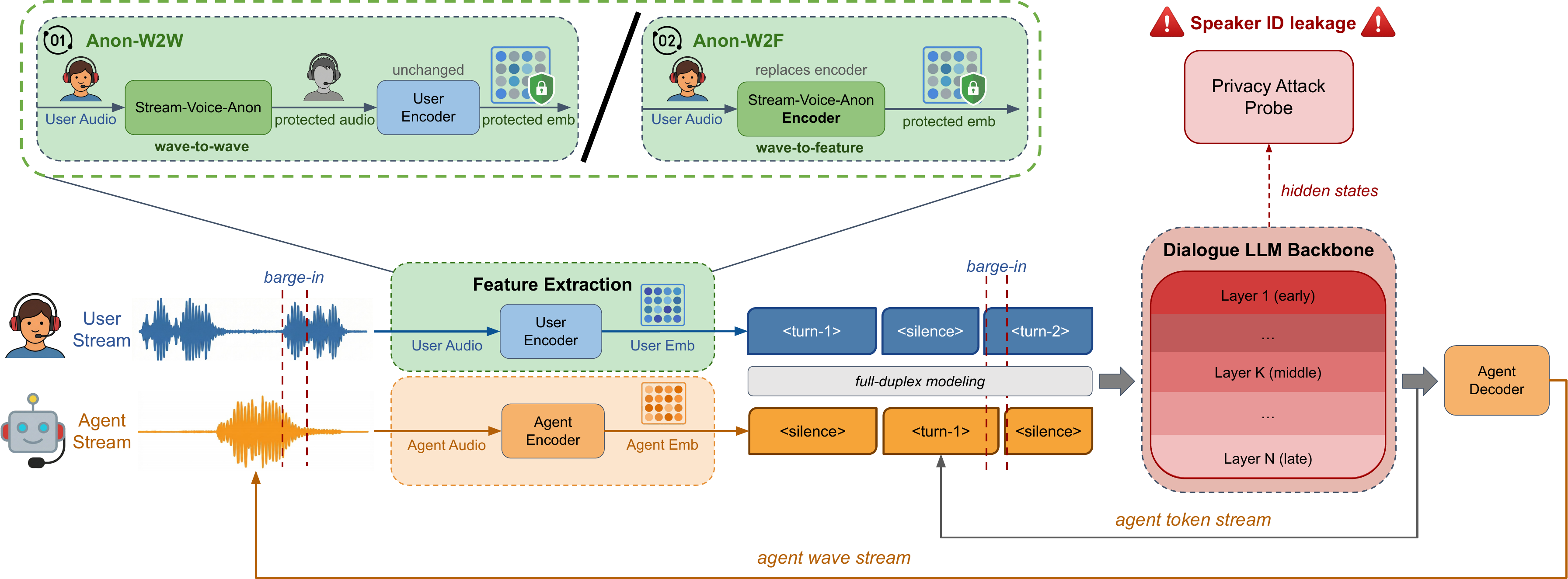}
  \caption{Overview of the original SALM-Duplex pipeline and proposed
    anonymization setups.
    The main diagram shows the ASR-based encoder baseline: an ECAPA-TDNN
    probe attached to the LLM's hidden states (red dashed path) reveals
    substantial speaker identity leakage.
    The Anon-W2W inset (upper left) prepends Stream-Voice-Anon to anonymize
    the waveform before the unchanged ASR encoder.
    The Anon-W2F inset (upper right) replaces the ASR encoder with the
    Stream-Voice-Anon encoder (anonymization active) and fine-tunes the LLM,
    eliminating the redundant waveform synthesis step.
    (Anon-W2F is demonstrated for SALM-Duplex; the Anon-W2W setup is
    additionally evaluated on Moshi.)}
  \label{fig:overview}
\end{figure*}

We answer this question empirically for two prominent E2E full-duplex systems,
SALM-Duplex and Moshi~\cite{defossez2024moshi}.
Following the VoicePrivacy~2024 Challenge~\cite{voiceprivacy2024} evaluation
protocol with a lazy-informed attacker scenario, we train a speaker verification
attacker (probe) on hidden state representations extracted from each layer group.
Equal error rate (EER) serves as our primary privacy metric; we additionally
report Linkability from the legally validated evaluation
framework of~\cite{vauquier2025legally}.
Our mean-pooled EER for the original SALM-Duplex (continuous encoder) is
28.5\%, revealing significant speaker identity leakage, while Moshi's discrete
encoder reaches 6.4\% EER (near-perfect identification).
We further propose and evaluate two streaming anonymization
setups (Figure~\ref{fig:overview}) using Stream-Voice-Anon~\cite{streamvoiceanon}
to mitigate this leakage without sacrificing dialogue utility.

Our contributions are:
\begin{itemize}
  \item We characterise speaker identity leakage in the hidden states of two
    E2E full-duplex dialogue LLMs (SALM-Duplex and Moshi), extending probing
    methodology~\cite{tenney2019bert, pasad2021layerwise} from static encoders
    to always-on dialogue backbones.
  \item We provide a layer-wise and turn-length-wise analysis showing which
    parts of the LLM carry the most speaker-identifying information and how
    leakage accumulates over dialogue length.
  \item We propose two streaming anonymization setups: Anon-W2W applies
    Stream-Voice-Anon at the waveform level (retaining the original encoder,
    validated on both SALM-Duplex and Moshi), while Anon-W2F replaces the
    continuous encoder with a discrete encoder and activates anonymization
    at the feature level.
\end{itemize}

\section{Related Work}

\subsection{E2E speech dialogue language models}
The generative spoken language model (GSLM)~\cite{lakhotia2021generative}
established the paradigm of modelling speech directly from discrete units,
subsequently extended to two-channel spoken dialogue by
dGSLM~\cite{nguyen2023generative}, the first fully E2E full-duplex dialogue
model.
SpeechGPT~\cite{zhang2023speechgpt} and SpiRit-LM~\cite{nguyen2024spiritlm}
demonstrated that a single LLM backbone can natively interleave speech and text
tokens for conversational interaction.
SyncLLM~\cite{emnlp2024synclm} introduced synchronised LLM inference to support
real-time full-duplex turn management.
The most recent systems, SALM-Duplex~\cite{salmduplex2025} and
Moshi~\cite{defossez2024moshi}, achieve low-latency always-on operation through
continuous dual-stream processing.
Crucially, because all of these models process raw user speech as input,
speaker identity information may be inadvertently encoded in their LLM hidden
representations, yet none has been analysed from a privacy perspective.

\subsection{Speaker anonymization}
The VoicePrivacy Challenge~\cite{tomashenko2024voiceprivacy, voiceprivacy2024}
established standardised evaluation protocols and benchmarks for speaker
anonymization, using the EER of an ASV system as the primary privacy metric;
Vauquier et al.~\cite{vauquier2025legally} recently extended this with a legally
validated framework incorporating Linkability metrics aligned
with GDPR requirements.
Anonymization approaches include x-vector-based
methods~\cite{meyer22_slt, tomashenko2024voiceprivacy},
perturbation~\cite{kuzmin24_spsc}, phonetic intermediate
representations~\cite{meyer22_interspeech}, neural audio codec language
models~\cite{panariello2024nac, yao24_spsc}, and disentangled speech
representations~\cite{champion22_interspeech, aloufi2020privacy}.
Streaming anonymization systems, including
Stream-Voice-Anon~\cite{streamvoiceanon}, TVTSyn~\cite{quamer2026tvtsyn},
DarkStream~\cite{darkstream2025}, and~\cite{quamer2024streaming}, extend these
methods to the real-time setting required by E2E full-duplex models; we select
Stream-Voice-Anon as our front-end owing to its competitive privacy--utility
trade-off and open-source
availability\footnote{\url{https://github.com/Plachtaa/StreamVoiceAnon}}.
All of these approaches operate at the \emph{waveform level}.
To the best of our knowledge, our work is the first to analyse and mitigate
speaker identity leakage \emph{inside} a full-duplex LLM's hidden states.

\begin{table*}[t]
  \caption{Privacy, dialogue quality, and efficiency on the VPC2024 evaluation set.
    Privacy: higher EER and lower Linkability (Lnk) = stronger privacy;
    EER\,=\,50\% = perfect anonymization.
    Linkability for no-anon rows uses the ``orig'' attacker;
    for anonymized rows, the ``lazy-informed'' attacker~\cite{vauquier2025legally}.
    Quality: higher sBLEU / sBERT = better.
    Efficiency: RTFx = 1/RTF; FRL = First Response Latency (s);
    TTSR = Turn-Taking Success Rate; Int.L.\ = Interruption Latency (s);
    ISR = Interruption Success Rate.
    \textbf{Bold} = best per column; \underline{underline} = second best.
    Shaded rows = anonymized conditions.}
  \label{tab:main}
  \centering
  \small
  \setlength{\tabcolsep}{4pt}
  \begin{tabular}{lcc @{\hskip 8pt} cc @{\hskip 8pt} cc cc @{\hskip 8pt} ccccc}
    \toprule
    \textbf{Model} & \textbf{User} & \textbf{Anonymization}
        & \multicolumn{2}{c@{\hskip 8pt}}{\textbf{Privacy}}
        & \multicolumn{4}{c@{\hskip 8pt}}{\textbf{Quality}}
        & \multicolumn{5}{c}{\textbf{Efficiency}} \\
    \cmidrule(lr){4-5} \cmidrule(lr){6-9} \cmidrule(lr){10-14}
    \textbf{Name} & \textbf{Encoder} & \textbf{Type}
      & EER$\uparrow$ & Lnk$\downarrow$
      & \multicolumn{2}{c}{sBLEU$\uparrow$}
      & \multicolumn{2}{c@{\hskip 8pt}}{sBERT$\uparrow$}
      & RTFx$\uparrow$ & FRL$\downarrow$
      & TTSR$\uparrow$ & Int.L.$\downarrow$ & ISR$\uparrow$ \\
    & & & & & S2T & S2S & S2T & S2S & & & & & \\
    \midrule
      & discrete & --  & 6.4  & 0.90 & \textbf{7.18} & \textbf{6.85} & \underline{0.50} & \underline{0.48} & 17  & \underline{0.50} & 0.85 & 1.24 & 0.45 \\
      \multirow{-2}{*}{Moshi}
      & \gc discrete & \gc W2W & \gc\underline{36.9} & \gc 0.35 & \gc 5.20 & \gc 5.00 & \gc 0.39 & \gc 0.38 & \gc 1.6  & \gc 0.72 & \gc 0.72 & \gc 1.31 & \gc 0.38 \\
    \midrule
      & discrete & --  & 11.2 & 0.79 & 4.18 & 3.15 & 0.45 & 0.29 & \underline{238} & \textbf{0.44} & 0.85 & 1.03 & 0.96 \\
      & \gc discrete & \gc W2F & \gc\textbf{41.0} & \gc\textbf{0.23} & \gc 3.57 & \gc 1.49 & \gc 0.39 & \gc 0.20 & \gc 2.5 & \gc\underline{0.50} & \gc 0.68 & \gc 1.18 & \gc 0.93 \\
    \cmidrule(l){2-14}
      & continuous & --  & 28.5 & 0.29 & \underline{6.91} & \underline{6.35} & \textbf{0.59} & \textbf{0.50} & \textbf{263} & 0.68 & \textbf{0.97} & \underline{0.60} & \textbf{0.99} \\
      \multirow{-4}{*}{\shortstack[l]{SALM-\\[-2pt]Duplex}}
      & \gc continuous & \gc W2W & \gc 34.6 & \gc\underline{0.24} & \gc 6.46 & \gc 5.49 & \gc\underline{0.55} & \gc 0.45 & \gc 1.7  & \gc 0.80 & \gc\underline{0.94} & \gc\textbf{0.57} & \gc\underline{0.98} \\
    \bottomrule
  \end{tabular}
\end{table*}

\subsection{Privacy in neural speech and language representations}
Probing studies have established that self-supervised speech models encode
substantial speaker identity across their transformer layers: Pasad
et al.~\cite{pasad2021layerwise} showed speaker information peaks in the lower
and middle layers of wav2vec~2.0, and SUPERB~\cite{yang2021superb} demonstrated
that lightweight probes achieve strong speaker verification from frozen SSL
representations.
Chiu et al.~\cite{chiu2025probing} recently confirmed this at large scale across
multiple SSL architectures and speaker attributes.
The layer-wise probing methodology itself was established by Tenney
et al.~\cite{tenney2019bert} for text models.
Our focus is distinct: rather than protecting conversational content, we
investigate \emph{speaker identity} leakage, the persistent encoding of
\emph{who is speaking}, which enables re-identification regardless of what is
said and may additionally extend to attributes such as gender, accent, and
health status.
Nautsch et al.~\cite{nautsch2019preserving} provide a comprehensive survey of
privacy threats in speaker and speech characterisation, identifying
re-identification from audio representations as a key risk category.

We extend this line of work to E2E full-duplex dialogue LLMs, whose hidden states, unlike static SSL encoders, are computed continuously over live user speech.

\section{Methodology}

\subsection{E2E full-duplex dialogue models}
We analyse two E2E full-duplex dialogue systems.
\textbf{Moshi}~\cite{defossez2024moshi} uses a decoder-only Transformer backbone
with a residual-quantization (RVQ) audio codec encoder to jointly model user and
system streams in a single autoregressive pass.
\textbf{SALM-Duplex}~\cite{salmduplex2025} is a decoder-only LLM that encodes
user audio via an ASR-initialized continuous encoder and processes dual
user--agent audio streams frame-synchronously, maintaining persistent hidden
states over the entire conversation; we reimplement a modified version that
adopts the speech decoder architecture from~\cite{wu2025chronological} but
without the chronological thinking mechanism.
Our reimplementation yields comparable response quality; the effect of the chronological thinking mechanism on privacy is left for future work.
Both systems maintain a persistent LLM hidden state over continuous user speech,
making them candidates for speaker identity probing.

\subsection{User encoder variants}
\label{sec:encoders}

We distinguish two encoder families by their output representation, which determines how much speaker identity reaches the LLM backbone.

Discrete encoders (Moshi and SALM-Duplex variant):
Moshi natively uses an RVQ codec encoder that produces discrete token
representations.
For SALM-Duplex we replace the continuous front-end with a discrete encoder
based on the Firefly architecture (implementation details in
Section~\ref{sec:setup}).
Because both discrete encoders are trained for high-fidelity speech
reconstruction, they preserve rich speaker information in their token
representations.
The SALM-Duplex discrete encoder can be operated in two modes: with
anonymization \emph{disabled}, serving as our ablation to isolate the effect of
the encoder swap alone; or with anonymization \emph{active} via
Stream-Voice-Anon~\cite{streamvoiceanon}, constituting our Anon-W2F
setup (Section~\ref{sec:anon_setups}).

Continuous encoder (original SALM-Duplex): An ASR-initialized
adapter encodes raw user audio into continuous embeddings.
Because ASR pretraining optimizes for speech content rather than speaker
characteristics, this encoder represents a good design choice for built-in
partial privacy protection, though with a limited privacy ceiling as we show
in Section~\ref{sec:results}.

\subsection{Anonymization setups}
\label{sec:anon_setups}

We propose two streaming anonymization setups using Stream-Voice-Anon~\cite{streamvoiceanon}.

\subsubsection{Anon-W2W: Wave-to-wave anonymization}
Anon-W2W pairs with the continuous encoder for SALM-Duplex and with
Moshi's native discrete codec encoder.
Stream-Voice-Anon is applied as a pre-processing step that transforms the raw
user waveform into an anonymized waveform before it is fed to the dialogue
model.
However, it introduces a redundant processing step: the anonymized waveform is
first synthesized and then re-encoded by the model's original encoder.

\subsubsection{Anon-W2F: Wave-to-feature anonymization}
Anon-W2F pairs with the discrete encoder (anonymization active).
We replace the continuous encoder front-end of SALM-Duplex with the discrete
encoder and enable Stream-Voice-Anon's anonymization module, which operates
natively on the discrete token representations.
Because the anonymization step operates in the same feature domain as the
encoder, this eliminates the redundant waveform synthesis step of Anon-W2W.
The modified model is pretrained on a similar data mixture as~\cite{wu2025chronological}
(approximately 12k hours of multi-turn dialogue and 2.7k hours of QA data),
then fine-tuned on
InstructS2S-200K\footnote{\url{https://huggingface.co/datasets/ICTNLP/InstructS2S-200K}}~\cite{fang2025llamaomni2}.
This setup requires architectural modification but achieves stronger privacy
guarantees through feature-domain-native anonymization.
While the Anon-W2F principle (replacing a model's encoder with an
anonymization-capable discrete encoder) is architecturally transferable, its
effectiveness depends on the target model's tokenizer compatibility; we
demonstrate it for SALM-Duplex and leave Moshi integration for future work.

\subsection{Hidden state extraction}
We extract hidden state representations from both Moshi ($N{=}32$ transformer
layers) and SALM-Duplex ($N{=}20$) at three individual layers: \textbf{early}
(layer~1), \textbf{mid} (layer~$N/2$), and \textbf{late} (layer~$N$), as well
as a \textbf{mean-pooled} representation averaging across all $N$ layers
(denoted ``All'' in Table~\ref{tab:layerwise}).
Temporal pooling is handled internally by the probe architecture.

\subsection{Setup}
\label{sec:setup}

\textbf{Dataset.} Privacy metrics are evaluated on the VoicePrivacy 2024
Challenge~\cite{voiceprivacy2024} evaluation set, derived from LibriSpeech
dev-clean and test-clean splits.
While this dataset consists of read speech rather than spontaneous conversation,
it serves as a good starting point given the standardised evaluation protocol;
evaluating on conversational corpora is an important direction for future work.
The speaker verification attacker is trained on LibriSpeech train-clean-360.
Quality and efficiency metrics are evaluated on
MtBenchEval~\cite{yan2025uro, lin2025full_v15}, a multi-turn dialogue benchmark
standardised by URO-Bench.
Moshi uses open-source pretrained weights; SALM-Duplex is reimplemented
following the speech decoder architecture of~\cite{wu2025chronological}.

\noindent\textbf{Speaker verification attacker.}
We train separate ECAPA-TDNN~\cite{desplanques20_interspeech} attackers from
scratch for SALM-Duplex and Moshi, replicating the lazy-informed attacker
scenario from VPC2024~\cite{voiceprivacy2024}: identical hyperparameters, with
an input projection layer adapted to each model's hidden state dimension.

\noindent\textbf{Privacy metrics.} Following the VoicePrivacy 2024
Challenge~\cite{voiceprivacy2024} protocol, we report equal error rate (EER\%)
as the primary privacy metric.
A higher EER indicates stronger privacy protection; EER $\approx 50\%$
corresponds to random-chance discrimination, i.e., complete anonymization.
To further strengthen privacy analysis, we additionally report
Linkability~\cite{vauquier2025legally}.

\noindent\textbf{Utility metrics.} Dialogue quality is assessed using evaluation scripts
adapted from URO-Bench~\cite{yan2025uro},
Full-Duplex-Bench~\cite{lin2025full_v15}, and
SALM-Duplex~\cite{salmduplex2025}.
Quality metrics (sBLEU, sBERT) are reported separately for speech-to-text (S2T)
and speech-to-speech (S2S) response modes: in S2T mode the model outputs text
directly; in S2S mode the model outputs speech which is subsequently transcribed
by an ASR system.
Efficiency metrics include Real-Time Factor Speedup (RTFx), First Response
Latency (FRL), Turn-Taking Success Rate (TTSR), Interruption Latency (Int.L.),
and Interruption Success Rate (ISR), reported per system condition
(Table~\ref{tab:main}).
All RTFx and latency measurements are estimated on a single GPU.

\begin{table}[t]
  \caption{Layer-wise EER (\%) of the speaker verification attacker.
    Higher = better privacy; 50\% = chance level.
    All = mean-pooled over all layers.
    \textbf{Bold} = best; \underline{underline} = second best.
    Shaded rows = anonymized conditions.}
  \label{tab:layerwise}
  \centering
  \small
  \setlength{\tabcolsep}{3pt}
  \begin{tabular}{lll ccc|c}
    \toprule
    \textbf{System} & \textbf{Encoder} & \textbf{Anon.}
      & \textbf{Early} & \textbf{Mid} & \textbf{Late} & \textbf{All} \\
    \midrule
      & discrete   & --  &  7.3 &  5.6 &  6.4 &  6.4 \\
      \multirow{-2}{*}{Moshi}
      & \gc discrete   & \gc W2W & \gc\underline{42.5} & \gc\underline{37.6} & \gc\underline{35.2} & \gc\underline{36.9} \\
    \midrule
      & discrete   & --  &  7.5 & 14.0 & 20.1 & 11.2 \\
      & \gc discrete   & \gc W2F & \gc\textbf{43.8} & \gc\textbf{40.5} & \gc\textbf{40.1} & \gc\textbf{41.0} \\
    \cmidrule(l){2-7}
      & continuous      & --  & 24.6 & 28.6 & 32.1 & 28.5 \\
      \multirow{-4}{*}{\shortstack[l]{SALM-\\[-2pt]Duplex}}
      & \gc continuous      & \gc W2W & \gc 31.5 & \gc 33.7 & \gc 35.3 & \gc 34.6 \\
    \bottomrule
  \end{tabular}
\end{table}

\section{Results}
\label{sec:results}

Table~\ref{tab:main} reports privacy, response quality, and efficiency metrics.
Discrete encoders, trained for high-fidelity speech reconstruction, leak
substantially more speaker identity than the continuous encoder (Moshi: 6.4\%
EER; SALM-Duplex discrete: 11.2\% vs.\ continuous: 28.5\%), confirming that
ASR pretraining acts as a privacy-positive design choice by discarding speaker
characteristics in favour of linguistic content.
Both anonymization setups substantially reduce leakage: Anon-W2W raises EER to
36.9\% for Moshi (+30.5 points) and 34.6\% for SALM-Duplex continuous (+6.1
points), while Anon-W2F achieves 41.0\% EER, approaching the 50\% chance ceiling.
Crucially, comparing Anon-W2F against the discrete-only ablation (11.2\%) confirms
the gain is entirely from anonymization, not the encoder swap.
These patterns hold across both architectures, confirming speaker identity
exposure as a consistent property of E2E full-duplex LLM hidden states.
Our ECAPA-TDNN probe represents a \emph{lower bound} on leakage; the
VoicePrivacy Attacker Challenge~\cite{vpc_attacker2025} has shown that more
sophisticated attackers can further reduce EER, suggesting that actual privacy
risk may be higher than reported here.

Anonymization introduces moderate quality degradation (sBERT S2T drops 7--22\%
relative) but privacy gains consistently outweigh the cost, with EER improving
21--477\% relative across setups.
We note that our text-based quality metrics do not capture speech-level
attributes such as naturalness and prosody; speech quality evaluation (e.g.,
MOS, UTMOS) remains for future work.
Adding the anonymization module reduces RTFx substantially (from 17--263$\times$
to 1.6--2.5$\times$) as the anonymizer dominates inference time, but all
conditions remain real-time viable (RTFx $> 1$).
Anon-W2F is faster than Anon-W2W (RTFx 2.5 vs.\ 1.6--1.7) because the
feature-domain encoder avoids the redundant waveform synthesis step.
Reducing the anonymization module's computational cost is an important direction
for future work.

\subsection{Ablation studies}

\textbf{Layer-wise analysis.}
Table~\ref{tab:layerwise} shows per-layer EER.
Moshi exhibits uniformly low EER across all layers (5.6--7.3\%), while both
SALM-Duplex variants show decreasing leakage from early to late layers,
consistent with deeper layers progressively abstracting away speaker features.
Anonymization raises EER uniformly across all layer groups, with all anonymized
conditions reaching the 31--44\% range.

\noindent\textbf{Turn-length analysis.}
Figure~\ref{fig:turnwise_link} plots $1 - \mathrm{Linkability}$ vs.\ turn count.
Without anonymization, privacy degrades rapidly within the first few turns.
Both anonymized systems maintain acceptable protection even after 10 turns
(SALM-Duplex + W2F: 0.90$\to$0.78; Moshi + W2W: 0.83$\to$0.67).

\section{Conclusion}

We have shown that significant speaker identity leakage is a consistent property
of E2E full-duplex architectures, demonstrated across both SALM-Duplex and
Moshi, with discrete encoders leaking substantially more than the
ASR-pretrained continuous encoder.
Waveform-level anonymization (Anon-W2W) reduces this risk across
both architectures, with the largest relative gains observed for the most
exposed systems, while feature-domain anonymization (Anon-W2F) achieves the
strongest protection, raising EER by over 3.5x and approaching chance level.
These findings motivate privacy-by-design for always-on speech AI systems.

Future work includes extending Anon-W2F to Moshi and other E2E full-duplex
architectures, studying personalization with minimal privacy threats,
improving privacy protection systems to have less impact on response quality
and latency, and evaluating privacy under stronger and more diverse attacker
models.

\section{Generative AI Use Disclosure}
Generative AI tools were used for proofreading and grammar correction with minor
changes. All research ideas, experimental design, implementation, and analysis
were conducted by the authors.

\bibliographystyle{IEEEtran}
\bibliography{references}

\end{document}